\documentclass[aps,prl,superscriptaddress,twocolumn,floatfix,showpacs,longbibliography,nobibnotes]{revtex4-1}
\usepackage{graphicx}
\usepackage{overpic}
\usepackage{dcolumn}
\usepackage{bm}
\usepackage{hyperref}
\usepackage{amsfonts}
\usepackage{xcolor}
\usepackage{comment}
\usepackage{physics}
\usepackage{booktabs}
\usepackage[normalem]{ulem}
\usepackage{subfigure}
\usepackage{makecell}
\usepackage{braket}
\usepackage{float}

\usepackage{amsmath} 

\usepackage{ulem} 

\usepackage{dsfont}

\begin{document}

\title{Axion like particles multi-parameter sensing}

  \author{Hassan Manshouri}
  \email[]{h.manshouri@ph.iut.ac.ir}
  \affiliation{Department of Physics, Isfahan University of Technology, Isfahan 84156-83111, Iran}
  \affiliation{Quantum Technology Research Group, Isfahan University of Technology, Isfahan 84156-83111, Iran}
  
  \author{Moslem Zarei}
  \email[]{m.zarei@iut.ac.ir}
  \affiliation{Department of Physics, Isfahan University of Technology, Isfahan 84156-83111, Iran}
  \affiliation{Quantum Technology Research Group, Isfahan University of Technology, Isfahan 84156-83111, Iran}
   
  \author{Mehdi Abdi}
  \email[]{mehabdi@iut.ac.ir}
  \affiliation{Department of Physics, Isfahan University of Technology, Isfahan 84156-83111, Iran}
  \affiliation{Quantum Technology Research Group, Isfahan University of Technology, Isfahan 84156-83111, Iran}

   \date{\today}

\begin{abstract}
The search for the axion like particles (APLs)---one of the deepest puzzles in modern cosmology---may hold the key to understanding dark matter and dark energy.
In this work, we introduce a setup taking advantage of the quantum metrology techniques to constrain the hypothetical mass and coupling constants of APLs by employing exotic pseudoscalar spin-spin interactions between fermions mediated by ALPs.
A key advantage of our approach is the exploitation of position-dependent spin sensor results to the high sensitivity of the probe.
To simultaneously investigate the axion mass and coupling constants, we invoke a multi-probe detection strategy.
Through this strategy, we circumvent the singularity of the quantum Fisher information matrix as an ultimate upper bound on the sensitivity of any probe.
For the axion masses in the range of $m_a {\le} 10^{-3} \text{eV}$, this setup can exclude values of axion coupling constants $g^e_pg^n_p$ down to $10^{-7}$~.
\end{abstract}

\maketitle

\paragraph{Introduction---}

Axions are hypothetical particles originally proposed as a solution to the strong CP problem \cite{Wilczek1978,Peccei1977,Kim2010}.
Their mass and couplings to other particles have not yet been experimentally confirmed.
In theories at ultrahigh energies, axions and axion like particles (collectively referred to as axions) can interact with other intermediating particles in exotic pseudoscalar spin-spin interactions~\cite{Kim2010}.
Recent experiments, searching for exotic spin-spin interactions on broad range of particle masses, have been examined using advanced techniques and setups such as magnetometry~\cite{Kim2019,Crescini2022,QUAX2020,Hajebrahimi_2023,Zarei_2022,Sharifian_2023}, nuclear magnetic resonance~\cite{Jiang2021,Afach:2021pfd,Crescini2018}, torsion-pendulum measurements~\cite{Heckel2013}, trapped ions~\cite{Wineland1991}, nitrogen-vacancy centers in diamond~\cite{Rong2018,Chigusa:2024psk}, geo-electrons~\cite{Hunter2013}, neutron beams~\cite{Yan2013}, masers~\cite{Glenday2008}, cantilevers~\cite{Ding2020}
and even a network of probes~\cite{GNOME:2023rpz,Fukuda:2025zcf,Jiang:2025uga,Afach:2021pfd}. 

For enhancing the measurement precision of an unknown parameter, fundamental quantum advantages and features such as entanglement~\cite{giovannetti2004quantum}, squeezing~\cite{Grangier1987}, resonance~\cite{Dutykh2020,manshouri2026}, and criticality~\cite{Venuti2007,MONTENEGRO2025} has been harnessed. 
The fundamental bound for sensing an unknown single-parameter is theoretically given by the Cram\'er-Rao inequality, which provides a complete understanding of single-parameter sensing~\cite{fisher1922mathematical}.
This framework puts a tight bound on precision, and identifies the optimal measurement basis of this bound~\cite{MONTENEGRO2025,2009quantum}.
However, the simultaneous estimation of several unknown parameters needs a multi-parameter metrological strategy~\cite{Liu2020Quantum}.
While the Cram\'er-Rao formalism, in principle, captures the ultimate bound of multi-parameter precision, one needs to proceed with caution.
Indeed, the covariance matrix quantifying the precision is bounded by the inverse of Fisher information matrix (FIM), $\mathcal{F}$, using the Cram\'er-Rao bound (CRB) formalism~\cite{Liu2020Quantum}.
The precision can be tightened by optimizing over all possible measurement known as quantum Fisher information matrix (QFIM), $\mathcal{Q}$ \cite{Candeloro2024}.
As in this case the CRB is a matrix inequality, the bound for simultaneous estimation of multi-parameters often cannot be saturated as a result of incompatibility of optimal measurements for different parameters~\cite{Carollo2019}.
Another possible obstacle in simultaneous estimation of multiple parameters is the singularity of FIM~\cite{Goldberg2021}.
Although singularity of FIM is not a fundamental challenge~\cite{Yang2025}, the QFIM singularity is indeed fundamental as it gives rise to singular FIM for all possible measurement choices~\cite{Mihailescu2024}.

Stark localization, a phenomenon in condense matter physics, happens when the tunneling of particles in lattice to neighboring sites is suppressed due to existence of a gradient field~\cite{Wannier_1960}, including single- and multi-particle, without disorder~\cite{Guo2021,Manshouri_2025,LI2026}.  
It takes place at infinitesimal fields as the system size increases \cite{he2023stark}. Quantum enhanced sensitivity of Stark model can be a breakthrough in ultra-weak fields sensing. 
Using Stark model sensitive to external magnetic fields is theoretically a valuable sensing method for weak field sensing such as pseudomagnetic fields generated by the axion mediator.
However, in the case of axions, as both the axion mass and couplings with other particles have not been precisely measured yet, multi-parameter estimation techniques need to be employed. 
In this study, we introduce a setup and strategy to simultaneously provide a precision estimation of axion mass and axion pseudoscalar couplings using the QFIM and CRB. 
We overcome the singularity in the QFIM---which arises from the simultaneous estimation of both the axion mass $m_a$ and the coupling constants---by employing a multi-probe method.
To examine the performance of our method the axion exclusion plot is derived from our multi-probe setup.
The results show that it can surpass the current experiment precisions considering axion couplings $g^e_p g^n_p$ across a wider range of axion masses~\cite{Wang2022,Almasi2020}.

\paragraph{Multi-parameter estimation---}\label{sec:MPE}
For a properly prepared probe a vector of $d$ unknown parameters $\bm{\theta}= (\theta_1,\theta_2,...,\theta_d)^T$ can imprint the related information onto the dynamics of the probe state.
The precision with which one can estimate these parameters is given by a covariance matrix whose elements quantify the interdependence of estimating parameters.
The fundamental achievable precision of the multi-parameter estimation is set by the quantum CRB in the form of
\begin{equation}\label{eq:CRB}
	\text{Cov}[\hat{\bm{\theta}}] \geq \frac{1}{\mathcal{M} \mathcal{Q}}~,
\end{equation}
where $\mathcal{M}$ is the number of measurements, $\hat{\bm{\theta}}$ are the estimators of $\bm{\theta}$~,  and $\text{Cov}[\hat{\bm{\theta}}]$ is the covariance matrix with the elements
\begin{equation}
	\text{Cov}(\theta_a ~, \theta_b)= \big\langle\!\left(\theta_a-\braket{\theta_a}\right)\left(\theta_b-\braket{\theta_b}\right)\!\big\rangle.
\end{equation}
Accordingly, the QFIM elements for a pure state $\ket\Psi$ are given by~\cite{2009quantum}
\begin{equation} \label{eq:QFIM}
	\mathcal{Q}_{ab}= 4 \text{Re}[ \braket{\partial_{\theta_a}\Psi|\partial_{\theta_b}\Psi}-\braket{\partial_{\theta_a} \Psi|\Psi}\braket{\Psi|\partial_{\theta_b}\Psi}]~.
\end{equation}
In the case of single effective parameter, i.e. $h= f(\bm{\theta})$, one has $\ket{\Psi(\bm{\theta})}\equiv \ket{\Psi(h)}$, hence, all combinations of parameters that give the same value for $h$ result the same dynamics for the probe state~\cite{Mihailescu2024}.
On the level of quantum states of the system, this symmetry leads to a singular QFIM, namely $\det[\mathcal{Q}]=0$.
Indeed, using the chain rule $\ket{\partial_{\theta_a}\Psi}= \ket{\partial_{h}\Psi} \partial_{\theta_a} h$, the QFIM elements in \eqref{eq:QFIM} read
\begin{equation}\label{eq:QFIMh}
	\mathcal{Q}_{ab}= \mathcal{Q}_h \times (\partial_{\theta_a} h) (\partial_{\theta_b} h)~.
\end{equation}
It is straightforward to show that for two parameters $h{=}f(\theta_a , \theta_b)$ one has $\det [{\mathcal{Q}}]{=}0$.
This can be generalized to the case of $d\geq 3$~\cite{Mihailescu2024}.

A practical approach to resolve the singularity of the QFIM is to invoke the advantage of multi-probes with a state composed of tensor product of individual probe states.
According to properties of the QFIM for a tensor product state $\ket{\Psi(\bm{\theta})} {=} \bigotimes_n \ket{\Psi(\bm{\theta})}_n$, the QFIM equals to the sum over QFIM of the individual probes, $\mathcal{Q}(\bm{\theta}) {=} \sum_{n} \mathcal{Q}(\bm{\theta}_n)$ \cite{Liu2020Quantum,Toth2014}, thus, removing the singularity of the QFIM:
\begin{align}\label{eq:detQMul}
	\det [{\mathcal{Q}}] &= \sum_{nm} \mathcal{Q}_{h_n} \mathcal{Q}_{h_m}(\partial_{\theta_a} {h_n})^2 (\partial_{\theta_b} {h_m})^2 \\
	&\hspace{-3mm}- \sum_{nm} \mathcal{Q}_{h_n} \mathcal{Q}_{h_m} (\partial_{\theta_a} {h_n}) (\partial_{\theta_b} {h_m}) (\partial_{\theta_b} {h_n}) (\partial_{\theta_a} {h_m})
	\neq 0~.\nonumber
\end{align}
As a result of the non-singular QFIM, the simultaneous estimation of unknown parameters $\bm{\theta}$ becomes theoretically achievable.


\paragraph{Axion-mediated spin-spin interaction---}
Investigation of spin-dependent forces between macroscopic objects due to exchange of bosons between spin-1/2 fermions can result to the discovery of the fifth force~\cite{Cong2024}.
The Yukawa-like exotic spin-spin interaction mediated by axions, is written in the form of ($c=\hbar=1$)~\cite{Cong2024,Wang2022}
\begin{eqnarray}\label{eq:spin-spin-ex}
	V_{pp} = -\frac{g^e_p g^n_p}{4}
	\left[
	(\hat{\bm{\sigma}}_1\cdot\hat{\bm{\sigma}}_2)
	\left(\frac{m_a}{r^2} + \frac{1}{r^3}\right)
	\right. ~~~~~~~~~~~~~~~~\nonumber\\
	\left.
	-(\hat{\bm{\sigma}}_1\cdot\hat{\bm{r}})
	(\hat{\bm{\sigma}}_2\cdot\hat{\bm{r}})
	\left(\frac{m_a^2}{r} + \frac{3m_a}{r^2} + \frac{3}{r^3}\right)
	\right]
	\frac{e^{-m_a r}}{4\pi m_1 m_2}~,
\end{eqnarray}
where $g^e_p g^n_p$ is the product of electron and neutron pseudoscalar coupling constants \cite{Wang2022}, $\hat{\bm{\sigma}}_1$ ($\hat{\bm{\sigma}}_2$) is the spin vector of the sensor  (source) and $m_1$ ($m_2$) is the corresponding mass. $\bm{r}$ is the vector connecting the interacting spins, and $m_a$ is the axion mass.
Beside the spin dependency of Yukawa potential, it depends on  distance between the probe and spin source. This feature results to the probe's sensitivity to the spin sensor position. 

\paragraph{Probe configuration---}
\begin{figure}[t]
	\includegraphics[width=.49\linewidth]{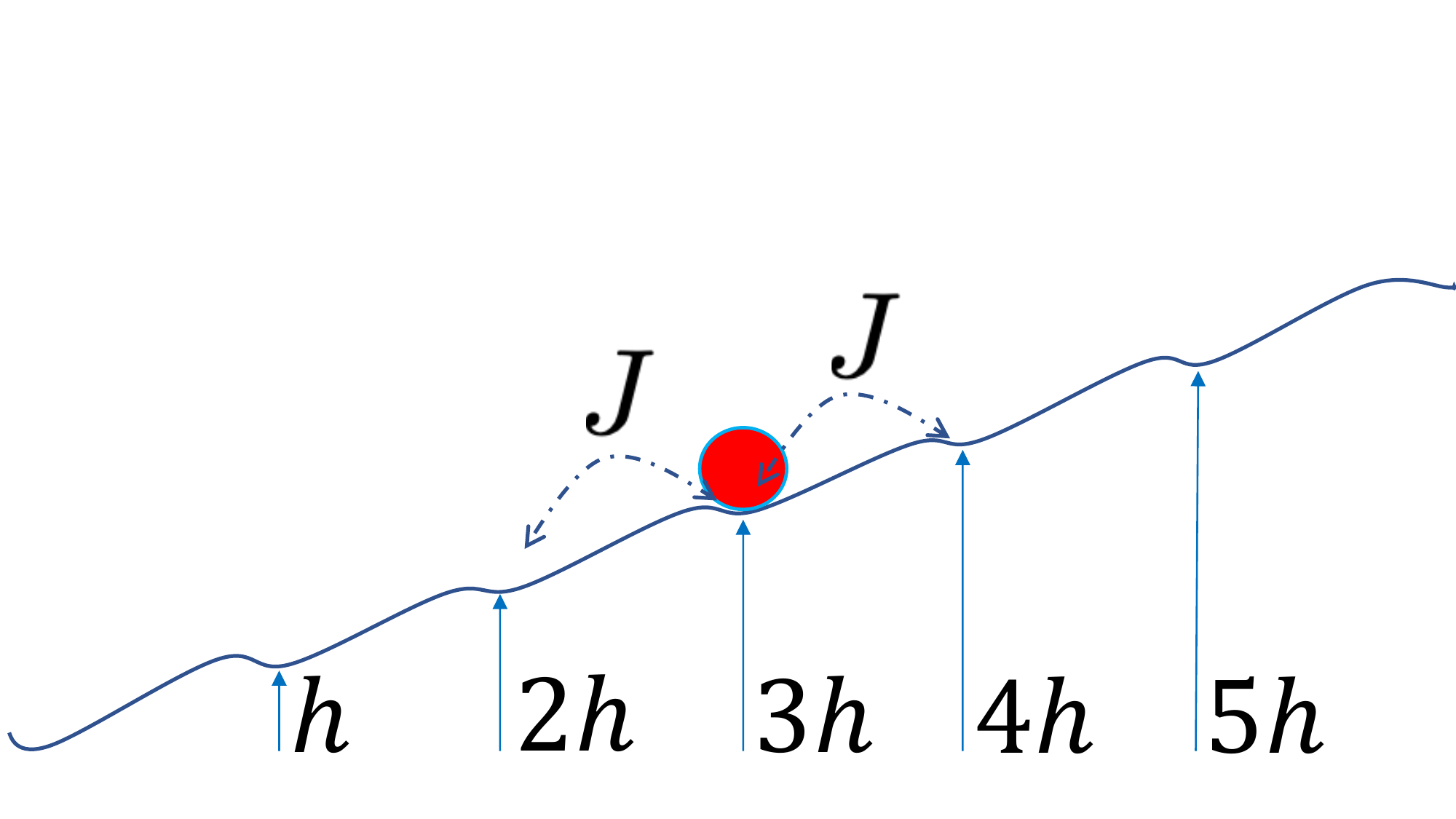}
	\put(-15,55){\color{black}(a)}
	\includegraphics[width=.49\linewidth]{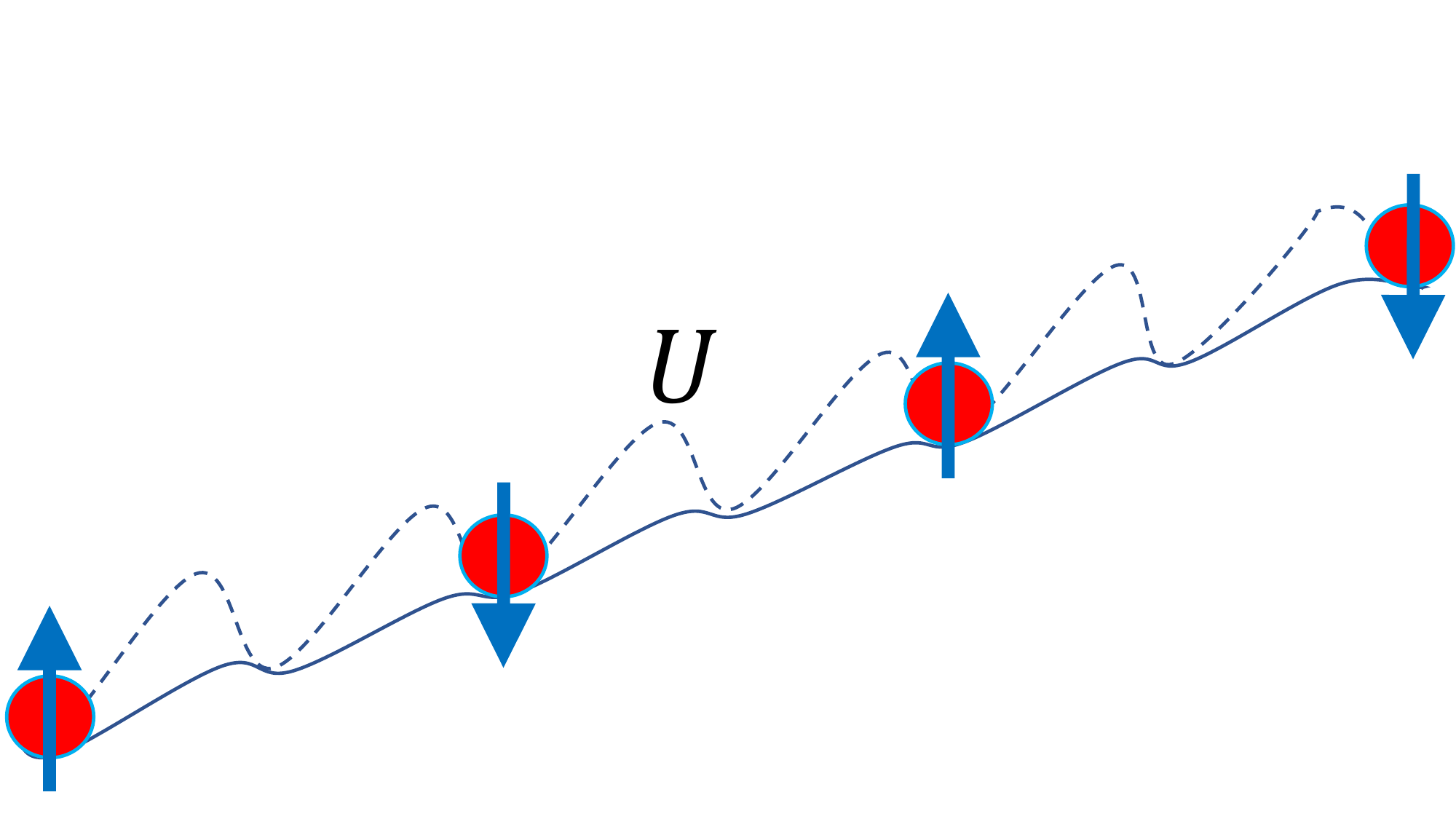}
	\put(-15,55){\color{black}(b)}
	\caption{(a) Schematic of the lattice with tunneling rate $J$ (dashed line) and gradient field $h$. (b) Half-filling configuration of probe spins, $\ket{. . . \uparrow0\downarrow0 . . . }$. Dashed line is the on-site interaction strength $U$ between spins.}
	\label{fig:lattice}
\end{figure}
For the probe we consider a lattice of interacting fermions in the presence of the gradient field, as illustrated in Fig. \ref{fig:lattice}.
We employ the Fermi-Hubbard model for the probe(s) whose Hamiltonian is given by \cite{Coleman2015}
\begin{equation}\label{eq:XXZhamiltonian}
	H = H_0 -J\sum_{l,\sigma} (\hat{a}^\dagger_{l+1,\sigma} \hat{a}_{l,\sigma}+\text{H.c.})
	+ U \sum_{l} \hat{n}_{l,\uparrow} \hat{n}_{l,\downarrow}~,
\end{equation}
where $H_0$ is the background field, $J$ denotes the tunneling rate, $\hat{a}_{l,\sigma}(\hat{a}^\dagger_{l,\sigma})$ the fermionic annihilation (creation) operator for spin $\sigma$ on site $l$, $U$ the on-site Hubbard interaction, and $\hat{n}_{l,\sigma}{=}\hat{a}^\dagger_{l,\sigma}\hat{a}^{}_{l,\sigma}$.
The background field $H_0$ corresponds to the interaction of fermionic spins in lattice with an external source of spins mediated by axions.
Given the distance, every particle in the probe experiences a different field in Eq.~\eqref{eq:XXZhamiltonian}. This gradient field depends on unknown parameters, interaction couplings $g^e_p g^n_p$ and axion mass $m_a$ as well as the orientation of the probe with respect to the source.

\begin{figure}[t]
	\hfill
	\begin{overpic}[width=.85\linewidth]{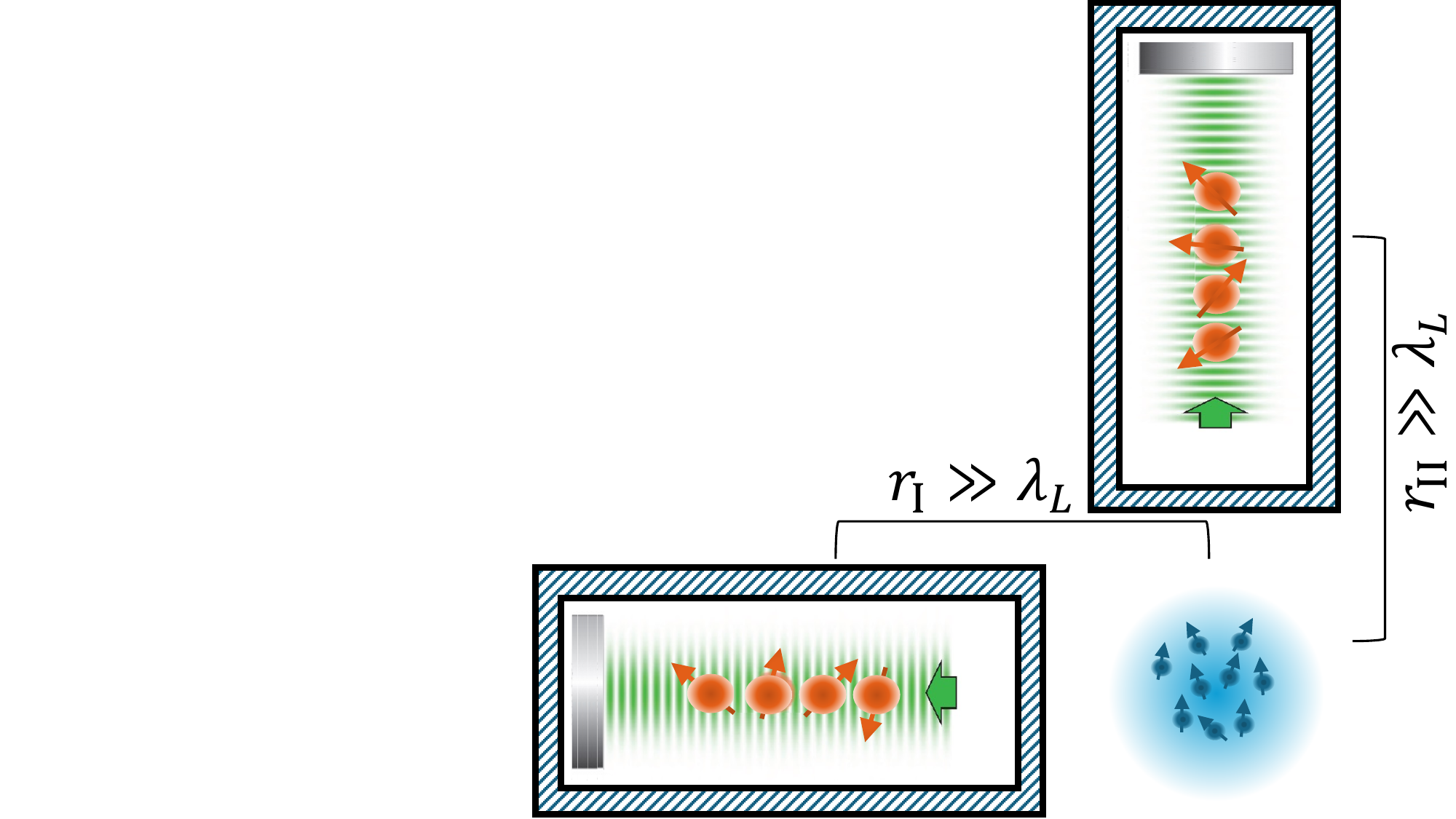}
		\put(90,60){\color{black}(b)}
		\put(-20,23){\includegraphics[width=.65\linewidth]{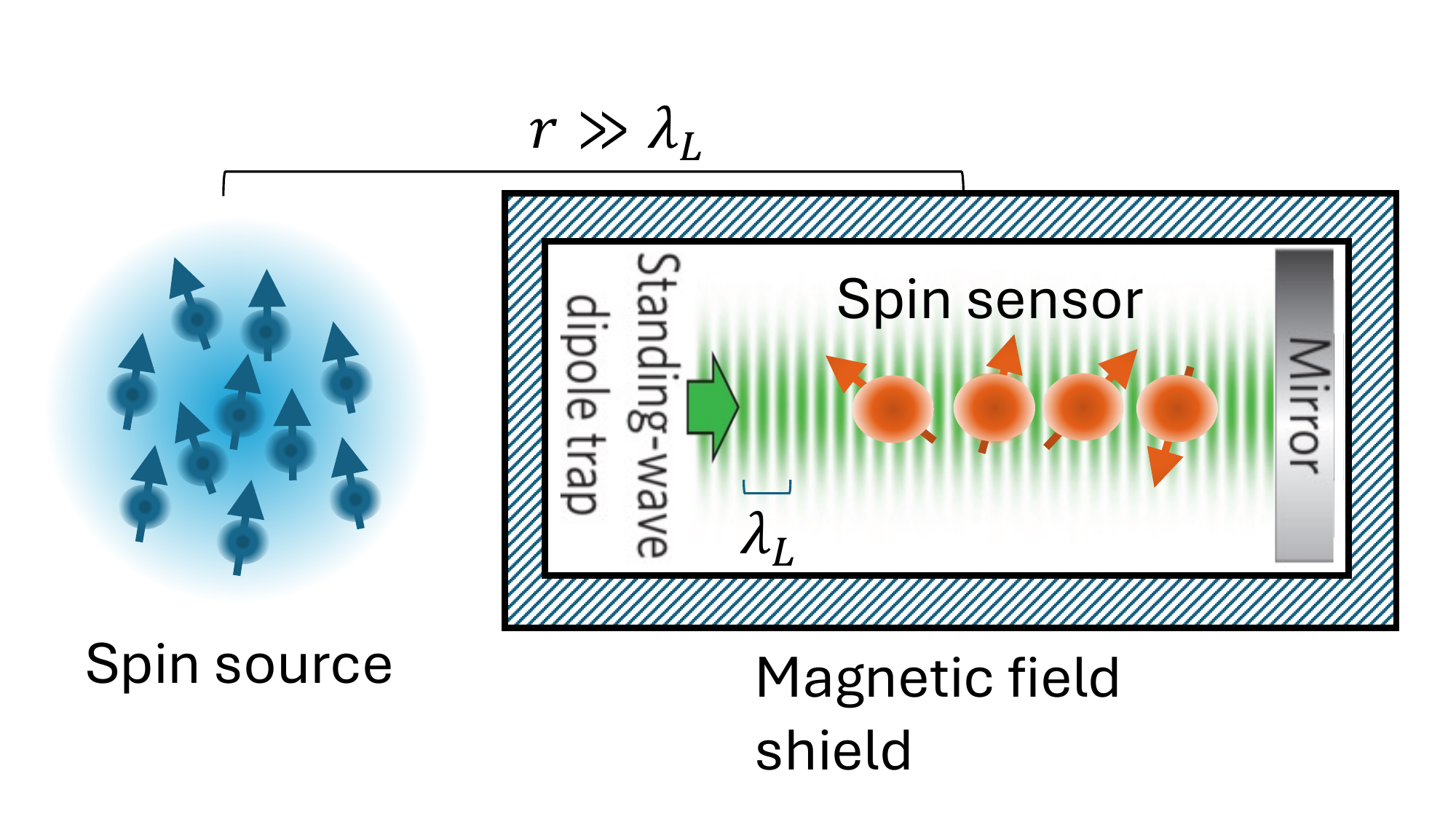}}
		\put(52,60){\color{black}(a)}
	\end{overpic}
	
	\caption{Schematic of sensing multi-parameter gradient field induced by axion. (a) Spins of source (blue spins) interacting with spins of sensor (orange) through exotic pseudoscalar spin-spin interaction. (b) Spin-source is positioned between two orthogonal probes to sense multi-parameters of axion field, $g^e_p g^n_p$ and $m_a$. The spin-source is positioned from the spin sensors by $r_{\text{I}}=r_{\text{II}}= r$~.}
	\label{fig:probe}
\end{figure}

As a possible experimental implementation, we consider neutrons of $^{3}\!\text{He}$ as neutron spin source \cite{JOHNSON1995} aligned with the $z$-axis in the lab frame, and an optical lattice with trapped $^{40}\!\text{K}$ electron spins as the sensor \cite{Kohlert2023fragmentation}. Spin sensors can interact with spin source due to the exchange of mediated axion, shown in Fig. \ref{fig:probe}(a). To measure the pseudomagnetic field of axion, the probe is magnetically shielded to eliminate the influence of the electromagnetic interaction with the source and any external magnetic field as the noise.
In our scenario the field gradient stems from the exotic pseudoscalar spin-spin interaction with the interaction in Eq. \eqref{eq:spin-spin-ex}.
To derive the effective gradient field we first consider an orientation where the source generates a transverse gradient field to the probe, see Fig.~\ref{fig:probe}(a).
Then we Taylor expand the equation \eqref{eq:spin-spin-ex} in terms of $r+\delta r$ and keep to the first order in $\delta r$, where $\delta r= l \lambda_L/2 $ with the lattice site number $l$.
Here, $\lambda_L$ is the wavelength of the optical trap forming the lattice.
By integrating over all polarized spin source $n$, the gradient field reads
\begin{equation}
	h_\perp (g^e_pg^n_p,m_a){\approx} \frac{\lambda_L}{2}\frac{g^e_p g^n_p}{4}M_2\left(\frac{m_a^2}{r^2}{+}\frac{3 m_a}{r^3}{+}\frac{3}{r^4}\right)
	\frac{e^{-m_a r}}{4 \pi  m_1 m_2}~,
\end{equation}
where $M_2 \equiv \sum_{k=1}^n \ev*{\hat\sigma_{2,k}^z}$~.
Using the Jordan-Wigner transformation \cite{Supp} for the corresponding spin operator of the probe $\hat{\sigma}^z_1$, one retains the gradient background field in Eq.~\eqref{eq:XXZhamiltonian} as $H_0 {=} h_\perp \sum_{l,\sigma} l \hat{n}_{l,\sigma}$~.

\paragraph{Sensitivity of  the probe---}
The sensitivity of a probe can determine using QFIM information in the form of
\begin{equation}
	\text{Tr}[\text{Cov}(\theta_1,\theta_2)] \geq \frac{\text{Tr}[\mathcal{Q}^{-1}]}{\mathcal{M} }~.
\end{equation}
Any singularity in the QFIM results to unbounded sensitivity of unknown parameters, meaning that the probe can not estimate them simultaneously even with an optimal estimator.

Under standard conditions, Bayesian estimation is used to obtain asymptotically optimal estimators in the large data set \cite{Berger1985,Escher_2011}. For a given set of $\mathcal{M}$ measurements, the distribution of encoded observed measurement data is determined by Bayes' theorem \cite{Berger1985}
\begin{equation}
	P\big(\bm{\theta}|\{n_k\}\big)= \frac{P\left(\{n_k\}|\bm{\theta}\right)P(\bm{\theta})}{P\left(\{n_k\}\right)},
\end{equation}
where $P\big(\bm{\theta}| \{n_k\}\big)$ is the Posterior conditional distribution of the parameter $\bm{\theta}$ based on the observed measurement data $\{n_k\}$.
In optical lattices, different configuration of particles are observed measurement data.
To ensure the normalization, the denominator $P(\{n_k\})$ is used to represent a valid probability distribution  \cite{Supp}. The likelihood function is
\begin{equation}
	P({n_k}|\bm{\theta}) = \frac{\mathcal{M}}{\prod_k n_k !} \prod_k [p^{th}_k(\bm{\theta})]^{n_k},
\end{equation}
where  $p^{th}_k(\bm{\theta})$ represents the conditional theoretical probability of the outcome $k$ given the parameters with the value $\bm{\theta}$ from a set of positive operator-valued measurements (POVMs), $\{\hat{\pi}_k\}$ \cite{2009quantum,Liu2020Quantum,MONTENEGRO2025}.

\begin{figure}[b]
	\includegraphics[width=.49\linewidth]{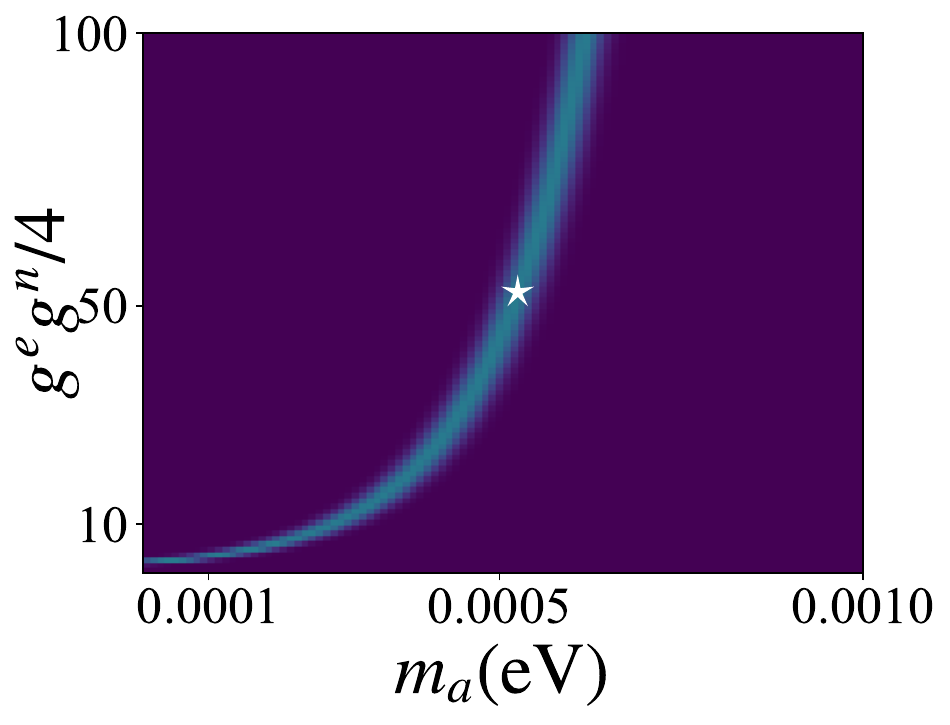}
	\put(-100,80){\color{white}(a)}
	\includegraphics[width=.49\linewidth]{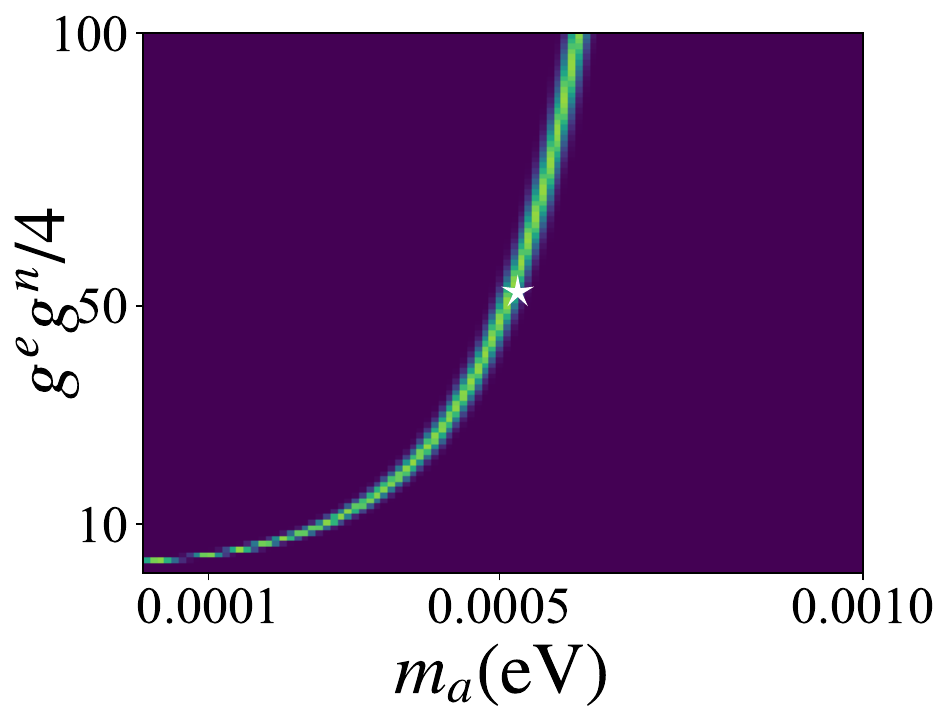}
	\put(-100,80){\color{white}(b)}\\
	\includegraphics[width=.49\linewidth]{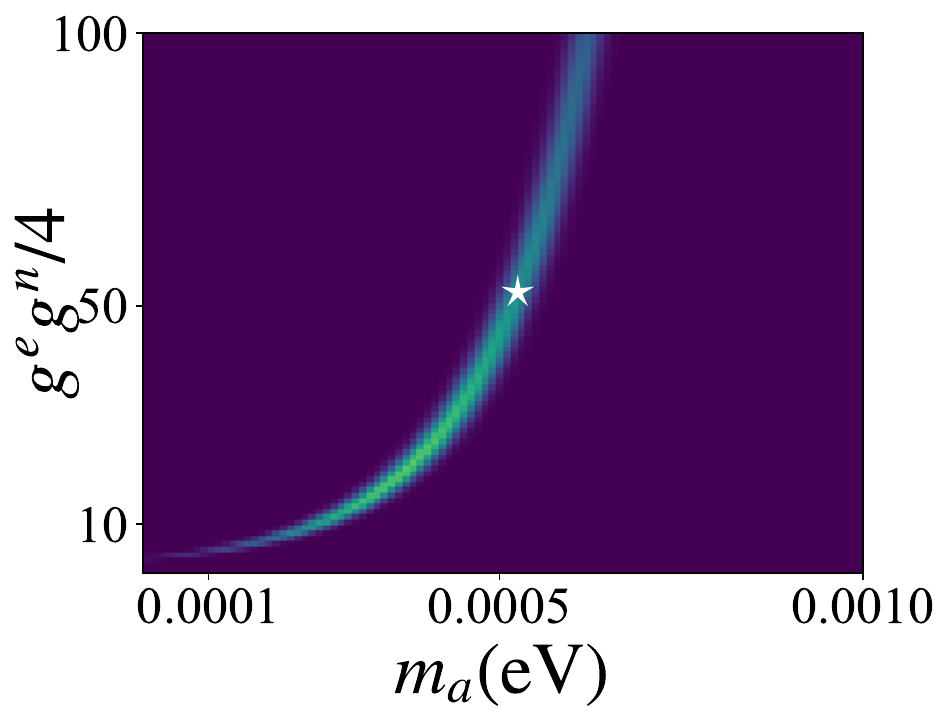}
	\put(-100,80){\color{white}(c)}
	\includegraphics[width=.49\linewidth]{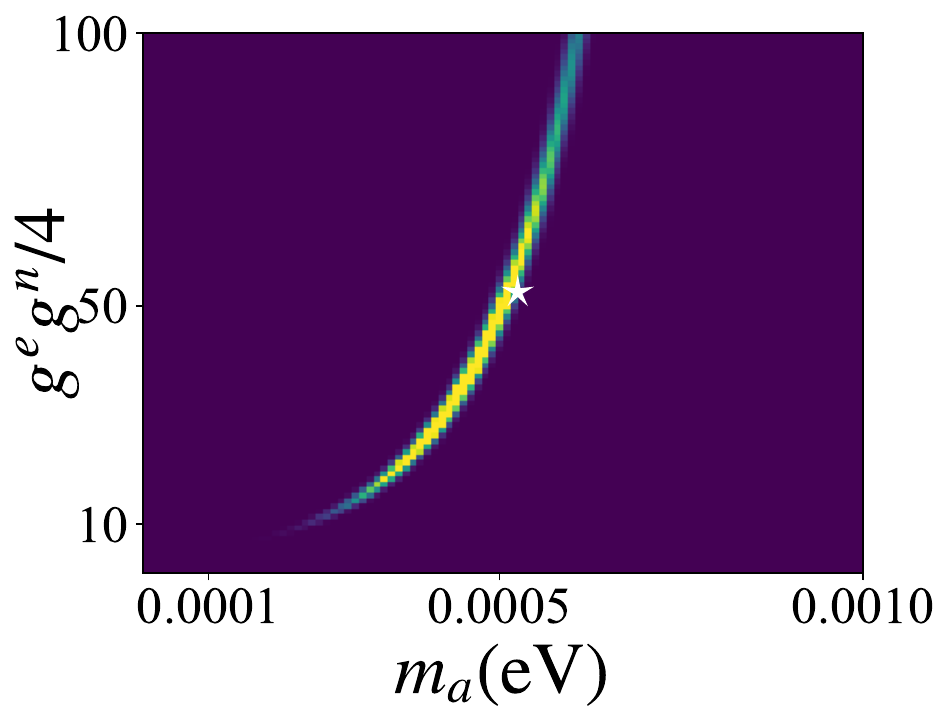}
	\put(-100,80){\color{white}(d)}
	\caption{Posterior probability distribution $P((g^eg^n,m)|n_k)$ of two unknown parameters $(g^eg^n,m_a)$ for system size $L=8$ with the half-filled particle numbers. For the single probe shown in Fig. \ref{fig:probe}(a), $P((g^eg^n,m)|n_k)$ is plotted with the number of measurement: (a) $\mathcal{M}=50$, and (b) $\mathcal{M}{=}200$. The corresponding  $P((g^eg^n,m)|n_k)$ of the two orthogonal probes configuration, shown in Fig. \ref{fig:probe}(b)(multi-probes), with the number of measurement: (c) $\mathcal{M}=50$, and (d) $\mathcal{M}=200$ is plotted. The star displays the true value of parameters, $g^e_p g^n_p/4= 50$ and $m_a= 5{\times} 10^{-4}$eV. 
		Colors show the amplitude of the probability. In all plots $J=4.13\times 10^{-13}$ eV~(100 Hz)~, $U=|J|$, and $n=10^{19}$ is the number of spin source with the prefect polarization.}
	\label{fig:Postgm}
\end{figure}
Using the probe shown in Fig. \ref{fig:probe}(a), posterior of two unknown parameters $(\theta_1,\theta_2){=} (g^e_p g^n_p , m_a)$ is plotted in Fig. \ref{fig:Postgm}(a) and (b)  \cite{Supp}.
The singularity in $\mathcal{Q}^{-1}$ results to a region with uniform posterior probability [Fig.~\ref{fig:Postgm}(a)], showing non-accessibility of simultaneous estimation of two unknown parameters.
Even by increasing the number of measurement $\mathcal{M}$ the singularity is sustained, notice the equal probabilities along the bright line in Fig.~\ref{fig:Postgm}(b).
In the following we utilize a second probe to resolve this singularity.

\paragraph{Resolving the singularity of the QFIM---}\label{sec:MPE}
As proved in Eq. \eqref{eq:detQMul}, using multi-probes can eliminate the singularity of the QFIM. 
Therefore, we employ a second probe with the gradient field aligned longitudinally with respect to its sites, see Fig.~\ref{fig:probe}(b).
Hence, one has
\begin{equation}
	h_\parallel {\approx} - \frac{\lambda_L}{2}\frac{g^e_p g^n_p}{4}M_2\left(\frac{m_a^3}{r} {+} \frac{3 m^2_a}{r^2}{+}\frac{6 m_a}{r^3}{+}\frac{6}{r^4}\right)
	\frac{e^{-m_a r}}{4\pi  m_1 m_2} ,
\end{equation} 
where we have assumed both probes are equally distanced from the source.
In this case, the background field, is aligned longitudinally. As a result, the first term in the Fermi-Hubbard Hamiltonian in Eq.~\eqref{eq:XXZhamiltonian} reads $H_0{=} h_\parallel \sum_{l} l (\hat{a}^\dagger_{l,\uparrow} \hat{a}_{l,\downarrow}+a^\dagger_{l,\downarrow} \hat{a}_{l,\uparrow}$)~ \cite{Supp}.
Using Bayesian estimation, the Posterior probability distribution converges to a point following the removal of the QFIM singularity.
Notice the local probability maximum in Figs.~\ref{fig:Postgm}(c) and (d).
Despite the intrinsic singularity of $P((g^eg^n,m)|n_k)$, with increasing $\mathcal{M}$, the posterior of the multi-probes converges to true values of unknown parameters.

\paragraph{Determining the axion exclusion region---}
\begin{figure}[t]
	\includegraphics[width=.49\linewidth]{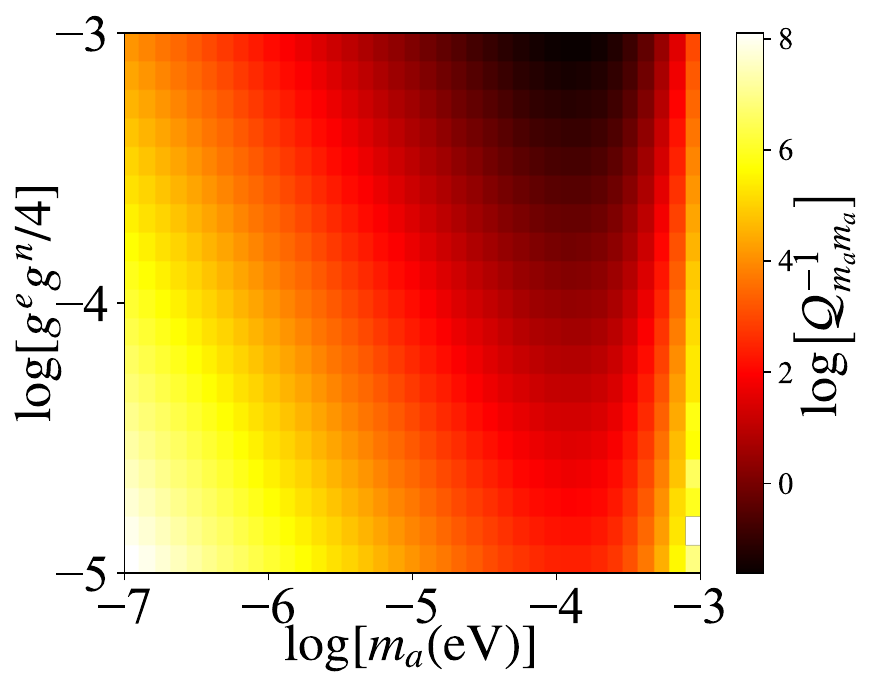}
	\put(-100,80){\color{black}(a)}
	\includegraphics[width=.49\linewidth]{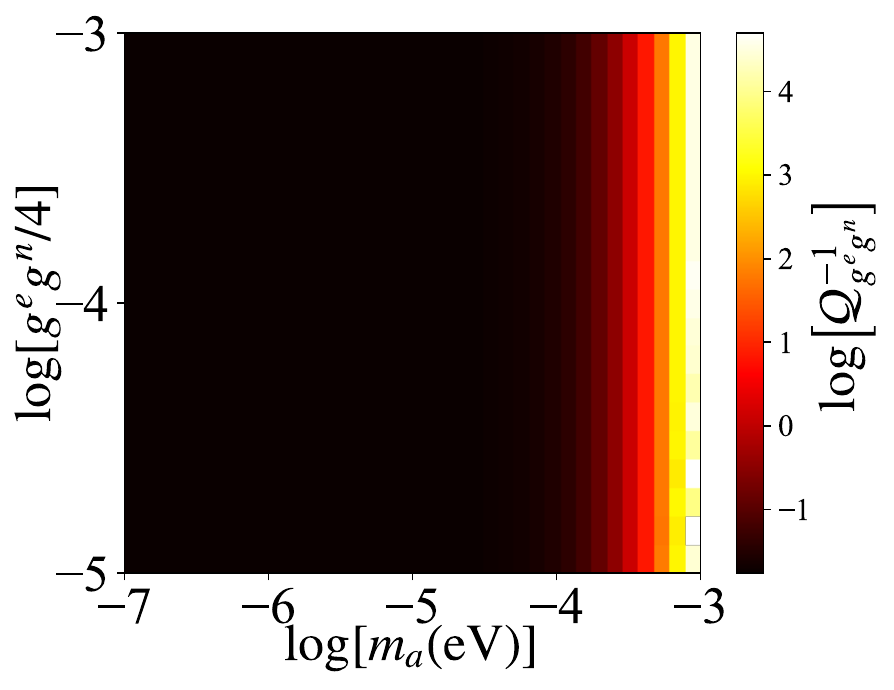}
	\put(-100,80){\color{black}(b)}
	\caption{(a)The simultaneous estimation of the QFIM inverse elements for the system size $L=8$~ with the half-filled particle numbers. For different values of axion mass $m_a$ and pseudoscalar couplings $g^e_pg^n_p$ in the logarithmic values, the corresponding inverse of the QFIM elements (a) $\mathcal{Q}^{-1}_{m_am_a}$ and (b) $\mathcal{Q}^{-1}_{g^eg^n}$ are plotted. In all plots, the tunneling rate is fixed to $J=4.13\times 10^{-13}$ eV~(100 Hz)~, the repulsive on-site interaction is $|U|=J$, and $n=10^{19}$.}
	\label{fig:QFIM}
\end{figure}
Determining the sensitivity of the axion multi-probes regardless of any estimator, can be estimated using inverse of the QFIM  and the CRB in Eq.~\eqref{eq:CRB}. Calculating diagonal elements of $\mathcal{Q}^{-1}_{mm}$ and $\mathcal{Q}^{-1}_{gg}$ results to ultimate sensitivity of simultaneous estimation of unknown parameters, $\text{Var}(m_a){\geq} \sqrt{\mathcal{Q}^{-1}_{mm}}/\sqrt{\mathcal{M}}$ and $\text{Var}(g^e g^n/4){\geq} \sqrt{\mathcal{Q}^{-1}_{gg}}/\sqrt{\mathcal{M}}$, respectively.
The axion multi-probes is prepared in the ground state to achieve the highest QFI values for sensing axion mass and pseudoscalar coupling $g^e_p g^n_p$. 
Preparing ground states of fermionic atoms in optical lattices can be achieved through traditional adiabatic protocols or advanced methods such as single-site control and measurement of tunneling elements and chemical potentials \cite{Tabares2025}. This preparation process takes a few milliseconds for fermionic optical lattices \cite{berto2019,Liu2011}.

In Fig. \ref{fig:QFIM} the associated QFIM inverse, $\mathcal{Q}^{-1}_{mm}$ and $\mathcal{Q}^{-1}_{gg}$, for different values of axion mass and pseudoscalar couplings $g^e_p g^n_p$ are plotted. 
 \begin{figure}[t]
 	\includegraphics[width=.95\linewidth]{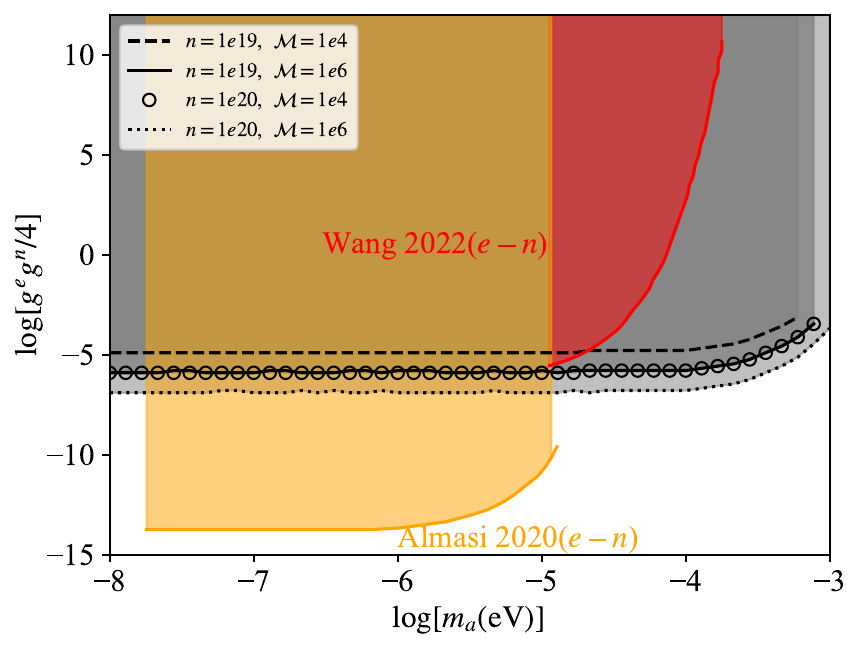}
 	\caption{ The exclusion region of axion mass and couplings under the simulation of  axion multi-probes using simultaneous estimation bounded by the Cram\'er Rao bound with the confidence level CL$=95\%$~. The multi-probes have been simulated with sensor system size $L{=}100$ with  the number of measurement repetitions $\mathcal{M}{=}10^4$ and $\mathcal{M}{=}10^6$~. The number of polarized spin source of $^{3}\!\text{He}$ is $n$ \cite{JOHNSON1995} with the distance $r= 1\text{cm}$~ from the sensor. The result of this work (black line) is compared with the current laboratory constraints on the exotic pseudoscalar interaction 
 	$g^e_p g^n_p$ between the studied particles pairs, such as \cite{Wang2022}(red line) and \cite{Almasi2020}  (orange line).
 	}
 	\label{fig:Excl}
 \end{figure}
 The exclusion region of the axion multi-probes parameters $(g_\text{excl}, m_\text{excl})$ where $g_\text{excl}\equiv (g^eg^n)_\text{excl}$~, can be connected to the QFIM in the form of
 \begin{equation}
 	g_\text{excl}(m) \simeq z_\alpha \sqrt{\mathcal{Q}^{-1}_{g^eg^n}(m)/\mathcal{M}}~,
 \end{equation}
  where $z_\alpha$ is the signal to noise ratio with the confidence level $\alpha$~. For the confidence level $\alpha= 95 \%$, in Fig. \ref{fig:Excl} we plot the exclusion region using the axion multi-probes for different polarized spin source $n$ and measurement repetition $\mathcal{M}$ for the system $L=100$ with the half-filled particle number. We have used the MPS simulation performed by the Tensor Network Python (TeNPy) package \cite{Hauschild2018}.
The plot is limited by $m_a\leq 10^{-3} \text{eV}$ due to the distance between the sensor and source $r\geq 1\text{cm}$.
With increasing $n$, $\mathcal{M}$, and also $L$ we can touch to the lower regions of axion couplings $g^e_p g^n_p$.

\paragraph{Conclusion---}
In this paper, we have introduced a new experimental setup to constrain the axion mass and coupling constants regarding the exotic pseudoscalar spin-spin interaction. Our approach exploits the position-dependent nature of the pseudoscalar spin-spin potential. Although simultaneous estimation of axion mass and coupling constants leads to the QFIM singularity, we overcome this challenge through a multi-probe detection strategy.
Applying the quantum Cram\'er-Rao bound---a cornerstone of quantum metrology---we have determined the exclusion region for the axion mass and couplings.
We demonstrate that even for sensing the weak pseudomagnetic field generated by axion-mediated interactions, our setup is sufficiently sensitive to constrain axion couplings $g^e_pg^n_p$ down to $10^{-7}$ in the axion masses range $m_a {\le} 10^{-3} \text{eV}$ that has not yet been excluded by existing experiments.

\begin{acknowledgments}
	
\end{acknowledgments}

\bibliography{reference-axion}


\newpage
\onecolumngrid

\begin{center}
	\textbf{\large SUPPLEMENTAL MATERIAL:\\ Axion like particles multi-parameter sensing}
	
	\bigskip
	
	Hassan Manshouri,$^{1,2}$ Moslem Zarei,$^{1,2}$ Mehdi Abdi,$^{1,2}$ \\
	
	\medskip
	
	$^{1}${\small \em Department of Physics, Isfahan University of Technology, Isfahan 84156-83111, Iran} \\
	$^{2}${\small \em Quantum Technology Research Group, Isfahan University of Technology, Isfahan 84156-83111, Iran} \\
\end{center}

\subsection{Jordan-Wigner transformation}
For a single spin the up and down states can be thought as empty and singly occupied fermion states in the form of
\begin{equation}
	\ket{\uparrow}  \equiv f^\dagger \ket{0}, ~~~~~~ \ket{\downarrow} \equiv \ket{0}~.
\end{equation}
In the form of spin-raising and spin-lowering operators we have
\begin{equation}
	S^+\equiv f^\dagger =\left(\begin{array}{cc} 0& 1\\ 0 & 0 \end{array}\right), ~~~~~~ S^{-}\equiv f =\left(\begin{array}{cc} 0& 0\\ 1 & 0 \end{array}\right) .
\end{equation}
Accordingly $x$, $y$ and $z$ components can be written as
\begin{eqnarray}
	&S_z&= f^\dagger f -\frac{1}{2}~,\nonumber\\
	&S_x&= \frac{1}{2} (f^\dagger+ f)~,\nonumber\\
	&S_y&= \frac{1}{2} (f^\dagger- f)~.
\end{eqnarray}
Spin operators satisfy the commutator algebra, $\left[S_a, S_b\right]{=} i\epsilon_{abc} S_c$, while due to a hidden supersymmetry, they satisfy an anticommuting algebra, $\{S_a, S_b\}{=} \frac{1}{2} \delta_{ab}$~. The problem is that for more than one spin, independent spin operators commute but independent fermions anticommute. To fix this problem, Jordan and Wigner attached a phase factor called a string to the fermions \cite{JordanWigner1928,Coleman2015}~. 
Then the Jordan-Wigner representation of the spin operators for a chain of spins in one dimension at site $j$ is written as
\begin{equation}
	S^+_l= f^\dagger_l e^{i\phi_l}~,
\end{equation}
where $\phi_j$ is the phase operator summing over all fermions at sites to the left of $j$, and $\phi_j= \pi \sum_{l<j} n_j$~.
Then the Jordan-Wigner transformation for the $z$ component of spin operator is
\begin{equation}
	S^z_l= f^\dagger_l f_l -\frac{1}{2}~.
\end{equation}
In the real systems such as optical lattices, particles can occupy the lattice sites according to the fermionic principle rules as $\ket{0}, \ket{\uparrow}, \ket{\downarrow}$ and $\ket{\uparrow \downarrow}$~. So the spin operator represented as
\begin{equation}
	S_{z,l} \equiv \hat{a}^\dagger_{\sigma,l} \hat{a}_{\sigma,l}
\end{equation}
for the $z$ component where $\sigma$ is the spin of fermions such as $\uparrow$ and $\downarrow$~. 
Acting on the vacuum state, the operator $\hat{a}^\dagger_\uparrow$ ($\hat{a}^\dagger_\downarrow$) creates the fermionic spin $\ket{\uparrow}$ ($\ket{\downarrow}$). 
As the operator $S_x$ reverses the direction of spins, the $x$ component of spin operator can be written as
\begin{equation}
	S_{x,l}\equiv \hat{a}^\dagger_{l,\uparrow} \hat{a}_{l,\downarrow}+\hat{a}^\dagger_{l,\downarrow} \hat{a}_{l,\uparrow}~.
\end{equation}
This definition guarantees reversing of the fermionic spins of every particle in lattice sites.

\subsection{Posterior Distribution}
To find the Posterior conditional distribution of the parameter $\bm{\theta}$ based on observed measurement data $\{n_k\}$, $P\big(\bm{\theta}|\{n_k\}\big)$ in Eq. (12) we consider no prior information. So $P(\bm{\theta})$ is a uniform distribution over the interval of interest. For parameters $\theta_1$ and $\theta_2$ we have
\begin{equation}
	P(\bm{\theta})= \frac{\mathds{1}}{N_{\theta_1 }N_{\theta_2}}~,
\end{equation}
where $\mathds{1}$ is a matrix with the dimension $N_{\theta_1} \times N_{\theta_2}$, and $N_{\theta}$ is the number of values of the parameter $\theta$~. Each probe has its own likelihood, but in the Hilbert space of multi-probes we use
 \begin{equation}
 	P\big(\{n_k\}|\bm{\theta}\big)= P^{\text{I}}\big(\{n_k\}|\bm{\theta}\big) P^{\text{II}}\big(\{n_k\}|\bm{\theta}\big)~,
 \end{equation}
  that is a matrix with the dimension $N_{\theta_1} \times N_{\theta_2}$~.
Also, to ensure the normalization we sum over all  values of unknown parameters in the form of 
\begin{equation}
	P\big(\{n_k\}\big)= \sum_{\{\theta_1\}} \sum_{\{\theta_2\}} P\big(\{n_k\}|\bm{\theta}\big) P(\bm{\theta})~,
\end{equation}
where $\{\theta_1\}$ and $\{\theta_2\}$ are values of parameters $\theta_1$ and $\theta_2$~. This normalization results to the normalization of the posterior distribution $\sum_{\{\theta_1\}} \sum_{\{\theta_2\}} P\big(\bm{\theta}|\{n_k\}\big)= 1$~.

\end{document}